# Is the OWASP Top 10 list comprehensive enough for writing secure code?

Parth Sane
Rochester Insititute of Technology

pvs3325@rit.edu

*Abstract*— **The OWASP Top 10 is a list that is published by the Open Web Application Security Project (OWASP). The general purpose is to serve as a watchlist for bugs to avoid while writing code. This paper compares how many of those weakness as described in the top ten list are actually reported in vulnerabilities listed in the National Vulnerability Database (NVD). That way it makes it possible to empirically show whether the OWASP Top 10 list is comprehensive enough or not, for code weaknesses that have been found in the past decade.**

## I. INTRODUCTION

The Open Web Application Security Project (OWASP) publishes a well-known list of vulnerabilities ranked in order from one to ten. They aim to provide that list as a warning to developers to avoid creating vulnerabilities in the applications they write. On the other hand, the National Vulnerability Database is the datastore for all the vulnerabilities that have been formally reported and described. Each vulnerability usually has an assigned Common Vulnerabilities and Exposure (CVE) identification number. The CVE IDs usually have Common Weakness Enumeration (CWE) numbers which describe overall about the types of weakness observed within the reported CVEs. There are one to many relationships between CVEs and CWEs respectively.

## II. RESEARCH QUESTIONS

*A. What are the top 10 reported CWE over the past ten years in the National Vulnerability Database?*

This is a particularly interesting question to investigate because the national vulnerability contains a detailed record of all the reported vulnerabilities. If we can see the big picture to ascertain what the issues are with respect to reported security weakness in security vulnerabilities.

*B. What are the CWE IDs for vulnerabilities described by OWASP Top 10?*

This list will help us create an observed custom top 10 list for CWEs, which may be one of the most important results and observations for this paper. A ten-year sample period or snapshot should be robust in providing an overview on the kinds of weakness present within the software.

*C. Which is the highest reported CWE in the past ten years within the National Vulnerability Database and is it the same as OWASP Top 10 Rank 1?*

This will help us assess whether the recommendations of OWASP line in with what has been observed over all the reported vulnerabilities over the past ten years.

*D. How many vulnerabilities with respective CWEs in the National Vulnerability Database are actually present in OWASP Top 10 CWEs?*

This research question will help us determine whether the OWASP Top 10 list has served as a warning indicator reliably. Maybe, there is an interesting observation that is possible that software developers aren't paying enough attention to OWASP as an organization who are trying to spread awareness about the common security issues that software developers inadvertently introduce into their application. In general, OWASP is referring to web apps, but there is an overlap between web apps and non-web applications.

## III. DATA PREPARATION

The data feeds for the National Vulnerability Database (NVD) can be obtained in the JavaScript Object Notation format (JSON) [1]. One of the challenges with respect to analyzing how the vulnerabilities are reported is the fact that they are reported year by year and not a complete file. Additionally, they do not correspond to a table structure for easy analysis, since JSON is object oriented with nesting of JSON arrays within a JSON object. This results in a need to normalize the data into a tabular and flat format, so that it can be easily analyzed. A python script was created to help fix this problem since all online JSON to CSV file format converters were taking too long. This was performed by creating a custom python script so as to flatten the information through Pandas data frames into Series objects. Hence overall algorithm in this process involved are:

*A. Data Normalization*

Convert nested JSON objects into arrays and keep traversing until you get to the innermost level. This is performed by the normalize function from the 'pandas.io.json' package.

*B. Data Adaptation*

Convert resulting data frame into Series. The series enables us to inspect the data and then pick a particular column that we are interested in. If there is further nesting, go back to Step A once again.

## C. Data Extraction

Now extract column having CWEs for associated IDs for every vulnerability. Note: you may have more than one CWE number or ID for any given vulnerability.

## D. Data Conversion

Finally, we need to convert the data into an excel file that can be opened in a spreadsheet application and all results can be compiled.

```python
import pandas as pd
from pandas.io.json import json_normalize
import json
with open('nvdcve-1.1-2019.json') as data_file:
    data = json.load(data_file)
df = json_normalize(data, ['CVE_Items'])
cvecol = df['cve']
cvenormalized = json_normalize(cvecol)
problemtypedata = cvenormalized['problemtype.problemtype_data']

problemtypelist = problemtypedata.tolist()
testdf = pd.DataFrame(problemtypelist)
details = json_normalize(testdf[0], ['description'])
cvevalues = details['value']
cvedf = pd.DataFrame(cvevalues)

cvedf.to_excel(r'cve_counts.xlsx', sheet_name='2019', engine='xlsxwriter')
```

Fig. 1. Data Adaptation and Transformation Python code

Figure 1 gives us an exact idea about the algorithm implementation in Python 3.

The critical challenges that were faced while developing this were deciding the libraries like 'Pandas'[6] and 'json'. Furthermore, learning the functions in the application programming interface was a challenge and required a bit of trial and error.

The above algorithm implementation creates individual excel files with sheets that need to be merged manually in Microsoft Excel.

Furthermore, we use Microsoft Excel's pivot table functionality[5] to go ahead and compute the count of values for CWE IDs under all the tables from information obtained from the National Vulnerability Database.

## IV. RESULTS

Overall, there were some interesting observations. Let's try to answer them one by one according to the research question proposed research questions.

### A. What are the top 10 reported CWE over the past ten years in the National Vulnerability Database?

Overall, the top ten list for CWEs generated for the paper are:
1) CWE 119 – **Buffer Overflow**
2) CWE 79 – **Cross Site Scripting**
3) CWE 20 – **Improper Input Validation**
4) CWE 200 – **Information Exposure**
5) CWE 264 – **Permissions, Privileges, and Access Controls**
6) CWE 84 – **Improper Neutralization of Encoded URI Schemes in a Web Page**
7) CWE 310 – **Cryptographic Issues**
8) CWE 125 – **Out-Of-Bounds Read**
9) CWE 399 – **Resource Management Errors**
10) CWE 352 – **Cross-Site Request Forgery**

Arguably this list is the most significant contribution for the paper, with a comprehensive count and analysis for security weakness discovered in software over the past 10 years from 2010 to 2019. It primarily gives us a broad overview of the top 10 security weaknesses in the real world over the past decade.

### B. What are the CWE IDs for vulnerabilities described by OWASP Top 10?

OWASP Top 10 Describes CWEs ID under different categories[2]. We are not going into details as to what they are for brevity. They are:

a) Cat A1 Injection
   i) CWE-77 - Improper Neutralization of Special Elements used in a Command ('Command Injection')
   ii) CWE-78 - Improper Neutralization of Special Elements used in an OS Command ('OS Command Injection')
   iii) CWE-88 - Improper Neutralization of Argument Delimiters in a Command ('Argument Injection')
   iv) CWE-89 – Improper Neutralization of Special Elements used in an SQL Command ('SQL Injection')
   v) CWE-90 – Improper Neutralization of Special Elements used in an LDAP Query ('LDAP Injection')
   vi) CWE-91 – XML Injection (aka Blind XPath Injection)
   vii) CWE-564 – SQL Injection: Hibernate
   viii) CWE-917 – Improper Neutralization of Special Elements used in an Expression Language Statement ('Expression Language Injection')

ix) CWE-943 – Improper Neutralization of Special Elements in Data Query Logic

b) Cat A2 Broken Authentication
   i) CWE-287 - Improper Authentication
   ii) CWE-256 - Unprotected Storage of Credentials
   iii) CWE-308 - Use of Single-factor Authentication
   iv) CWE-384 - Session Fixation
   v) CWE-522 - Files or Directories Accessible to External Parties
   vi) CWE-523 - Unprotected Transport of Credentials
   vii) CWE-613 - Insufficient Session Expiration
   viii) CWE-620 - Unverified Password Change
   ix) CWE-640 - Weak Password Recovery Mechanism for Forgotten Password

c) Cat A3 Sensitive Data Exposure
   i) CWE-220 - Sensitive Data Under FTP Root
   ii) CWE-295 - Improper Certificate Validation
   iii) CWE-311 - Missing Encryption of Sensitive Data
   iv) CWE-312 - Cleartext Storage of Sensitive Information
   v) CWE-319 - Cleartext Transmission of Sensitive Information
   vi) CWE-320 - Key Management Errors
   vii) CWE-325 - Missing Required Cryptographic Step
   viii) CWE-326 - Inadequate Encryption Strength
   ix) CWE-327 - Use of a Broken or Risky Cryptographic Algorithm
   x) CWE-328 - Reversible One-Way Hash
   xi) CWE-359 - Exposure of Private Information ('Privacy Violation')

d) Cat A4 XML External Entities
   i) CWE-611 - Improper Restriction of XML External Entity Reference
   ii) CWE-776 - Improper Restriction of Recursive Entity References in DTDs ('XML Entity Expansion')

e) Cat A5 Broken Access Control
   i) CWE-22 - Improper Limitation of a Pathname to a Restricted Directory ('Path Traversal')
   ii) CWE-284 - Improper Access Control
   iii) CWE-285 - Improper Authorization
   iv) CWE-425 - Direct Request ('Forced Browsing')
   v) CWE-639 - Authorization Bypass Through User-Controlled Key

f) Cat A6 Security Misconfiguration
   i) CWE-16 - Configuration
   ii) CWE-209 - Information Exposure Through an Error Message
   iii) CWE-548 - Information Exposure Through Directory Listing

g) Cat A7 Cross Site Scripting
   i) CWE-79 - Improper Neutralization of Input During Web Page Generation ('Cross-site Scripting')

h) Cat A8 Insecure Deserialization
   i) CWE-502 - Deserialization of Untrusted Data

i) Cat A9 Using Components with known vulnerabilities

j) Cat A10 Insufficient Logging and Monitoring
   i) CWE-223 - Omission of Security-relevant Information
   ii) CWE-778 - Insufficient Logging

In the above list 'Cat A9' is specifically important. This refers to using software components that are known to be vulnerable. As such we cannot point towards a single CWE ID as it would point towards all the categories for that known CVE ID for the known vulnerable piece of software. The easiest mitigation for a security vulnerability is to ensure that patches are deployed.

All the CWEs under the above list correspond to different categories with varying degrees of severity. But they in total represent the weakness that OWASP warns developers to avoid in very much detail.

C. *Which is the highest reported CWE in the past ten years within the National Vulnerability Database and is it the same as OWASP Top 10 Rank 1?*

The highest reported CWE-119 (Buffer Overflow). It is observed that it is not the same as OWASP Top 10 Category A1 which is Injection. Injection is reported with CWE IDs 77,78,88,89,90,81,564,917 and 943 for the A1 category in OWASP Top 10. This can be visually observed in the bar chart in Fig. 2 given below.

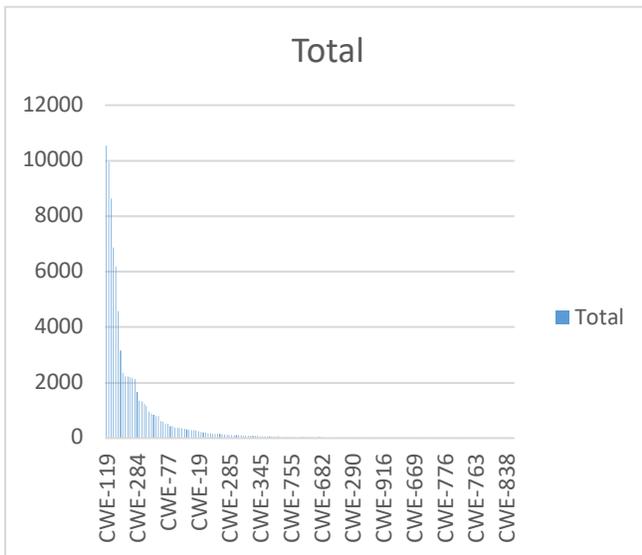

Fig. 2. Count of CWE IDs in the National Vulnerability Database for all vulnerabilites from 2010-2019.

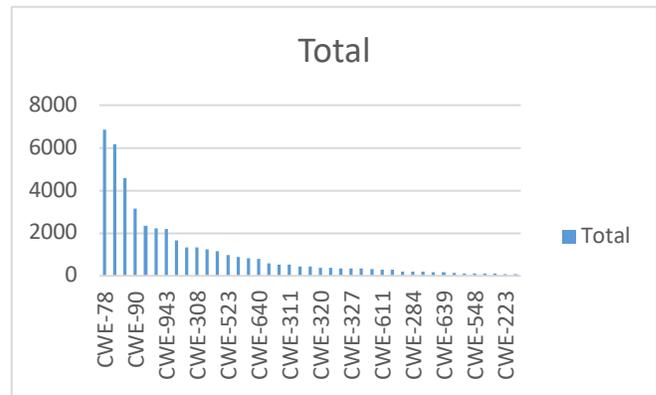

Fig. 3. Count of CWE IDs in the National Vulnerability Database for OWASP Top 10 vulnerabilites from 2010-2019.

Another interesting impact of Fig. 2 is to note that OWASP is really focusing on Web Application Security versus other application architectures like the desktop and mobile. But depending on the technology stack that is used to manipulate the data in the background, the malicious actors could bypass the security controls offered by the browser by default and then directly attack the underlying Application Programming Interface (API).

For example, consider a buffer overflow issue in the backend server code which is listening to HTTP requests. The attacker can specially craft a HTTP GET request[3] with parameters that may exceed these buffer limits and thus cause a buffer overflow with critical security consequences[7]. These may include arbitrary code execution possibly leading to complete compromise of underlying systems.

D. *How many vulnerabilities with respective CWEs in the National Vulnerability Database are actually present in OWASP Top 10 CWEs?*

The answer here seems to be pretty clear: The total number of weaknesses listed in the National Vulnerability Database are a superset of weakness listed in OWASP Top 10 by far. All of the listed weakness in OWASP Top 10 are included within the National Vulnerability Database. But this also means that the OWASP Top 10 list is not comprehensive enough, and developers should be aware of issues that may not be included within there.

Fig. 3 gives us an overview of the total count of CWE IDs that were observed for the duration of 2010-2019 in the National Vulnerability Database and were also present in the OWASP Top 10 list[2]. The OWASP Top 10 list has cross site scripting on 'Cat A7'. Assuming the list is a ranking of vulnerabilities to watch out for, we can observe that for a vulnerability that lies on the very top of the list calculated here in this paper from the results of the count of the National Vulnerability Database does not match with the top vulnerability described by the OWASP Top 10 list.

This is an interesting observation due to the fact that the OWASP Top 10 list has a different ranking as compared to the observed results for the past decade. Perhaps, there needs to be a better compiled list like described within this paper that can better serve the needs of developers. That way they are more aware about mistakes that other engineers commit most frequently.

## LIMITATIONS

This study does not consider the change of software architecture through the ages. Earlier we used to have more desktop-based applications, whereas now we are migrating towards mobile, web and cloud-based applications. This different outlook may give different results if we can control variables for different software architectures overall.

What we need to also consider is that this study does not single out software solutions or suites for having those vulnerabilities with their respective weaknesses. For example, we do not point out by saying that Adobe Flash or Silverlight is insecure, and we should not use that piece of software even though that may be the case[3].

Some automated tools even recommend software according to some fixed criteria[8]. But this paper is more of an observation and analysis style work, rather than a paper that presents a tool for performing work. Although, the provided python script can easily be extended to fit the requirements of any project that needs this sort of slice and dice analysis for data feeds from the National Vulnerability Database.

Another limitation is the fact that OWASP considers vulnerabilities only for web application projects. This may limit the usefulness of this study to only web applications with respect to comparisons between OWASP Top 10 and vulnerabilities in National Vulnerability Database. But nonetheless, there is significant overlap logically and conceptually that it makes it worth exploring and understanding the results.